\documentclass[aps,prl,twocolumn,showpacs,groupedaddress]{revtex4}
\usepackage{graphics,bm}
\usepackage{graphicx}
\usepackage{amsmath, amssymb, graphics}

\newcommand{\beq}{\begin{equation}}
\newcommand{\eeq}{\end{equation}}
\newcommand{\bqa}{\begin{eqnarray}}
\newcommand{\eqa}{\end{eqnarray}}
\def\gsim{\mathrel {\vcenter {\baselineskip 0pt \kern 0pt
\hbox{$>$} \kern 0pt \hbox{$\sim$} }}}
\def\lsim{\mathrel {\vcenter {\baselineskip 0pt \kern 0pt
\hbox{$<$} \kern 0pt \hbox{$\sim$} }}}

\protect

\begin{document}
\title{Formation of matter-wave soliton molecules}
\author{U. Al Khawaja$^1$ and H. T. C. Stoof$\,^2$}
\affiliation{ $^1$\it Physics Department, United Arab Emirates University, P.O. Box
17551, Al-Ain, United Arab Emirates.\\
$^2$Institute for Theoretical Physics, Utrecht University, Leuvenlaan 4,
3584 CE Utrecht, The Netherlands.}

\date{\today}

\begin{abstract}
We propose a method of forming matter-wave soliton molecules that
is inspired by the recent experiment of Dris {\it et al.}
\cite{randy}. In the proposed set-up we show that if two solitons
are initially prepared in phase and with a sufficiently small
separation and relative velocity, a bound pair will always form.
This is verified by direct numerical simulation of the
Gross-Pitaevskii equation and by the derivation of the exact
interaction energy of two solitons, which takes the form of a
Morse potential. This interaction potential depends not only on
the separation but also on the relative phase of the solitons and
is essential for an analytical treatment of a host of other
problems, such as the soliton gas and the Toda lattice of
solitons.
\end{abstract}


\maketitle

{\it Introduction. ---} Interactions between solitons have been
the subject of intense study in the past few decades
\cite{solov,gordon,anderson}. This was motivated both by the
fundamental physics involved and by the prospects of important
technological applications \cite{books}. At the fundamental level,
solitons show a dual nature by behaving as classical particles but
also exhibiting their wave nature when they, for instance, scatter
off each other. In particular, it was first pointed out by Karpman
and Solov'ev using perturbation theory \cite{solov}, by Gordon
using the exact two solitons solution \cite{gordon}, and by
Anderson and Lisak employing a variational approach
\cite{anderson}, that at large distances the force between two
solitons decays exponentially with their separation and is
proportional to the cosine of their phase difference. From the
applications point of view, this interaction between solitons is
actually a problem since it leads, when using optical solitons in
fibers as data carriers, to the destruction of the information
stored in a sequences of solitons.

The exciting possibility of forming so-called soliton molecules
gives yet another stimulus to the problem. Such molecules have
indeed recently been realized experimentally with optical solitons
by Stratmann {\it et al.}, who also measured the relative phase of
the two solitons and studied their binding mechanism
\cite{mitschke1}. Matter-wave soliton molecules have, however, not
yet been created experimentally. Theoretically, soliton molecules
have been studied previously in two-dimensional Bose-Einstein
condensates without \cite{perez} and with \cite{santos} dipolar
interactions. In the latter case an effective molecular
soliton-soliton potential was found, which results from the
dipolar nature of the condensate and interlayer effects.
Experimentally, bright solitons and soliton trains were observed
as the remnants of a collapsing condensate that was created by
using a Feshbach resonance to switch the interaction from
repulsive to attractive \cite{expts}. It was found that
neighboring solitons repel each other in spite of the fact that
the interatomic interactions are attractive. A variational
calculation showed that the repulsive force is caused by a $\pi$
phase difference between the solitons \cite{usama_randy}, in
agreement with known result for optical solitons
\cite{solov,gordon,anderson}. The conditions in these experiments
are thus not favorable for forming bound states of solitons.
Instead, if the two solitons start with a zero phase difference,
they attract each other and may ultimately form a molecule. Our
analysis of the soliton-soliton interaction energy shows that
while this is indeed a necessary condition, it is not sufficient
for soliton molecule formation. The initial separation and
relative velocity need to be sufficiently small such that the
interaction is not dramatically weakened by the exponential tail
and the kinetic energy of the relative motion does not
considerably exceed the soliton-soliton interaction energy.

An adjusted version of a recent experiment by Dris {\it et al.}
\cite{randy} is suggested here as a possible experimental method
of realizing soliton molecules in strongly elongated Bose-Einstein
condensates. In the experiment of Ref.~\cite{randy}, a trapped
bright soliton was launched onto a potential barrier at the center
of a harmonic trap. Here partial transmission and reflection
created two solitons, which then after half an oscillation in the
trap recombine again into a single one. We will show, however,
that a soliton molecule can be formed if both the harmonic
trapping potential and the potential barrier are switched off at
the time when the two solitons reach their classical turning
points. This guarantees the relative velocity to be zero.
Moreover, the two solitons are also guaranteed to have a zero
relative phase then, since they were created coherently from the
same soliton. The separation between the solitons prepared in this
way is controlled by the curvature of the harmonic trap. Thus, it
appears that all the conditions for soliton molecule formation can
be met in such an experimental set-up. In the following, we show
that this is indeed the case via direct numerical simulations of
the Gross-Pitaevskii equation. This is then followed by an
analytical analysis of the soliton-soliton interaction energy,
which also confirms the feasibility of our proposal. Moreover,
finding an accurate interaction potential for two solitons at an
arbitrary separation, is important not only for the present
problem of forming soliton molecules, but may also lay the grounds
for a host of other problems involving solitons interactions such
as the soliton Toda lattice \cite{toda} and the soliton gas
\cite{santosgas}.

{\it Soliton molecule formation. ---} The evolution of solitons in
the original experiment of Ref.~\cite{randy}, as well as in our
modified proposal, can be described by the effectively
one-dimensional Gross-Pitaevskii equation (GPE)
\begin{equation}
\left[i {\partial\over\partial
t}+{1\over2}{\partial^2\over\partial z^2}- V_{\rm ext}(z)
+g\,|\Psi(z,t)|^2\right]\Psi(z,t)=0 \label{gp},
\end{equation}
where, here and throughout, lengths are scaled to the
characteristic length $a_z=\sqrt{\hbar/m\omega_z}$ of the harmonic
potential $m\omega_z^2z^2/2$, time to $1/\omega_z$, and the
wavefunction $\Psi(z,t)$ to $1/\sqrt{2a_z\omega_\perp/\omega_z}$,
with $\omega_z$ and $\omega_\perp$ the axial and radial
frequencies of the strongly elongated ($\omega_\perp\gg\omega_z$)
trapping potential, respectively. In these units, the strength of
the interatomic interaction will be given by the ratio $g=-a/a_z$,
where $a$ is the $s$-wave scattering length. A square or gaussian
potential barrier is added to the harmonic potential at the center
that is included in $V_{\rm ext}(z)$.

\begin{figure}[htb]
\includegraphics[width=6cm]{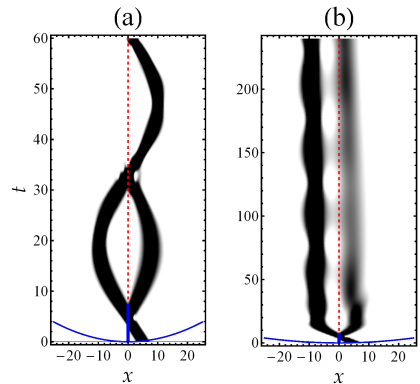}
\caption{(Color online) Spacio-temporal density plots for (a) the experiment of
Ref.~\cite{randy} and (b) our proposal for soliton molecule formation. In (a) a
single soliton is split by the potential barrier into two solitons that upon
returning to the trap center recombine into the original one. In (b) the
harmonic trapping potential and the potential barrier, shown with the solid
blue curve, are switched off at $t=15.75$ when the two solitons reach the
classical turning point. Initial soliton parameters: velocity $=-0.8$,
amplitude $=0.8$, position $=5$, and interaction strength $g=0.0113$. Square
barrier's parameters: height $=0.75$ and width $=0.5$.} \label{fig1}
\end{figure}

We now show that Eq.~(\ref{gp}) indeed accounts for the
interesting recombination of the two returning solitons at the
center, as can be seen in Fig.~\ref{fig1}(a). Repeating this
simulation for the case that we switch off the harmonic trapping
potential and the potential barrier at the point when the two
solitons reach their classical turning points, results in the
formation of a molecule of two solitons, as shown clearly in
Fig.~\ref{fig1}(b). The relative intensities or densities of the
two solitons in the molecule are set by the initial velocity of
the single original soliton and the height of the potential
barrier.

In another version of this experiment, one can also think of phase
imprinting one of the two solitons once it reaches the classical
turning point. In such an experiment, the phase of the cycle
performed by the molecule is controlled. For instance, one can
imprint a $\pi$ phase difference leading initially to an increase
of the separation of the solitons. If their initial separation is
small enough, however, they will ultimately return back and form
the molecule since their relative phase will change from $\pi$ to
zero as they move away from each other and the force becomes
attractive, as will be shown next.

{\it Soliton-soliton interaction energy. ---} As mentioned in the introduction,
the interaction energy of two solitons depends on their separation and on their
relative phase. It is also known that such an interaction may lead to the
formation of a bound state of two solitons. The key point here is that
dynamically the relative phase of two solitons changes as their separation
changes. If two solitons, that are initially in phase, approach each other
sufficiently slowly, their relative phase changes such that the force between
them becomes repulsive. When the two solitons then move away from each other,
the force becomes attractive again and the cycle completes itself. Thus, the
manner by which the relative phase changes with the separation between the
solitons is essential for understanding their binding properties. This fact is
qualitatively understood in the literature, but has never before quantitatively
been captured in a formula that accurately gives the interaction potential of
the solitons for both short as well as large separations. Here, we arrive at
this objective by deriving the exact dependence of the phase, and hence the
interaction potential, on the solitons separation. In addition to being of
importance by itself, this potential provides a convenient analytical
understanding for the possibility of soliton molecule formation.

Using the Inverse Scattering Method, the exact two soliton
solution of the homogeneous version of Eq.~(\ref{gp}) is derived
in Ref.~\cite{gordon}. In Ref.~\cite{usama_pre}, a re-derivation
puts the solution in terms of 8 parameters, namely the two initial
center-of-mass positions $z_{1,2}$ of the solitons, their
center-of-mass velocities $v_{1,2}$, their phases $\Phi_{1,2}$,
and their square amplitudes $n_{1,2}$. The separation between the
solitons can then be extracted analytically, from which also the
force follows in terms of these parameters. Here, we extend this
formalism to derive the interaction energy and to show that it
takes the form of a Morse potential.

For simplicity, the relative velocity is set to zero and the limit
of nearly equal soliton densities $n_-=n_2-n_1\ll n_+=n_2+n_1$ is
taken. The separation $\Delta(t)$ between the solitons and their
relative phase $\Phi(t)$ turn out to be given by
\begin{equation}
\Delta(t) =\frac{4}{\alpha}\,{\log\left[{
\alpha\left(\alpha^2+2\alpha\,\beta \cos \left(\omega\,t +\Phi
   _0\right)+\beta ^2\right)\over\beta\,g^2
n_-^2}\right]}\label{eq1},
\end{equation}
and
\begin{eqnarray}
\Phi(t)&=&\tan ^{-1}\left(\frac{\beta \sin \left(\omega\,t
+\phi_0\right)}{\alpha+\beta \cos \left(\omega\,t +\phi
_0\right)}\right)\nonumber\\&&- \tan ^{-1}\left(\frac{\alpha \sin
\left(\omega\,t +\phi_0\right)}{\beta+\alpha \cos \left(\omega\,t
+\phi _0\right) }\right)\label{eq2},
\end{eqnarray}
where $\beta=\exp{(\alpha\,x_0/4)}$, $\omega=\alpha\,gn_-/8$, and
$\alpha=gn_+$. In addition, $x_0$ and $\phi_0$ are two constants
that can be determined from the initial conditions
$\Delta(0)=\Delta_0$ and $\Phi(0)=\Phi_0$. The equations of motion
for $\Delta$ and $\Phi$ can now be obtained by differentiating
twice with respect to $t$. The resulting equations are simplified
by inverting Eqs.~(\ref{eq1}) and (\ref{eq2}) to express
$\cos{(\omega\,t+\phi_0)}$ and $\sin{(\omega\,t+\phi_0)}$ in terms
of $\Delta$ and $\Phi$. Finally, we obtain in this manner
\begin{equation}
{\ddot\Delta}=-{\alpha^3\over8}\,e^{-{1\over4}\alpha\,\Delta}\,\cos{\Phi}\label{eq100},
\end{equation}
\begin{equation}
{\ddot{\Phi}}={\alpha^4\over32}\,e^{-{1\over4}\alpha\,\Delta}\,\sin{\Phi}\label{eq200},
\end{equation}
which are identical to Gordon's formulae \cite{gordon}. Here,
however, we have the exact solutions of these equations of motion,
namely Eqs.~(\ref{eq1}) and (\ref{eq2}). This allows for obtaining
the force $\ddot\Delta$ in terms of $\Delta$ by solving
Eq.~(\ref{eq1}) for $\cos{(\omega\,t+\phi_0)}$ and substituting
the solution in Eq.~(\ref{eq100}). Integrating the resulting
expression with respect to $\Delta$ leads to the soliton-soliton
potential energy $V_{ss}=-\int {\ddot\Delta}\, d\Delta$
\begin{equation}
V_{ss}=\frac{({\alpha^2}/{4})\cos
{\Phi_0}}{e^{\frac{1}{4}\alpha{\Delta_0}}-e^{\frac{1}{4}
\alpha{\Delta_{\rm eq}}}}\left(e^{-\frac{1}{2} \alpha
   (\Delta -\Delta_{\rm eq})}-2\,e^{-\frac{1}{4} \alpha (\Delta
-{\Delta_{\rm eq}})}\right)\label{eq12},
\end{equation}
which has the form of a Morse potential with an equilibrium
position at $\Delta_{\rm eq}$ given by
\begin{equation}
\Delta_{\rm eq}=\Delta_0+\frac{4}{\alpha}\,{ \log
\left(\frac{\alpha^2-2 \alpha\,\beta\,\cos{\Phi_0}+\beta
^2}{\alpha^2+\beta
   ^2}\right)}\label{eq11}.
\end{equation}
Here, $\beta=\beta(\Delta_0,\Phi_0)$ corresponds to the real
solution of the initial conditions.

\begin{figure}[htb]
\includegraphics[width=6cm]{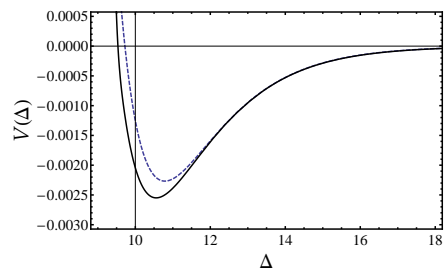}
\caption{Interaction potential between two bright solitons in
terms of their separation. Solid curve is obtained numerically
from the exact two solitons solution of the Gross-Pitaevskii
equation. Dashed curve corresponds to formula (\ref{eq12}).
Parameters: $n_-=0.2, n_+=4.7, g=0.5, \Phi_0=0$.} \label{fig2}
\end{figure}

Alternatively, the potential can be determined from the exact
solution directly. To that end, the location of the two solitons
are calculated numerically from the density $|\Psi(z,t)|^2$, from
which the solitons separation and the associated acceleration are
computed. Integrating the resulting force numerically, we again
obtain an interaction potential. In Fig.~\ref{fig2}, we plot the
interaction potential obtained by the two methods. It is clear
from this figure that the numerics captures the main features of
the potential in Eq.~(\ref{eq12}) apart from a slight change in
the location of the equilibrium point and depth of the potential.
The change is due to the fact that the difference in the
center-of-mass positions is numerically difficult to define
uniquely and is therefore not exactly equal to the parameter
$\Delta$ in the exact solution when the two solitons are too close
together and interference effects take place. The shape of this
potential, specifically the existence of a global minimum, proves
that a bound state of two solitons can be formed provided the
initial separation and relative phase are such that the energy is
negative.

\begin{figure}[htb]
\includegraphics[width=6cm]{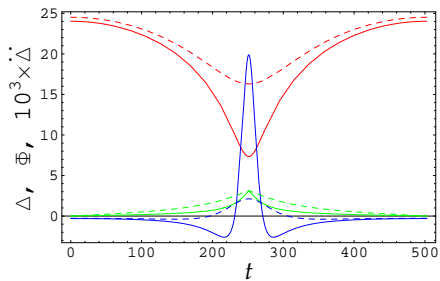}
\caption{(Color online) Two upper red curves: solitons separation
$\Delta(t)$. Two middle green curves: phase difference $\Phi(t)$.
Two lower blue curves: force ${\ddot\Delta}(t)$. Time range
corresponds to one period of the molecule's oscillation. Two
values of the initial separation were used: $\Delta_0=24.0$ (solid
curves) and $\Delta_0=24.5$ (dashed curves). Parameters:
$\Phi_0=0$, $n_-=0.1, n_+=1, g=1$.} \label{fig3}
\end{figure}

Finally, we notice that both the equilibrium point and depth of the potential
depend on the initial conditions. For further clarification of this rather
unusual situation, we plot in Fig.~\ref{fig3} the solitons separation
$\Delta(t)$ and relative phase $\Phi(t)$ during one period of the molecule's
oscillation. It is clear that by increasing the initial separation, the
equilibrium point around which the solitons oscillates also increases. Notice
that, for both cases, the phase changes from $0$ to $\pi$ over half the period
$\pi/\omega$ of the molecule's oscillation, which means that the force changes
from attractive to repulsive over the same period of time. On the other hand,
the dynamics of $\Delta(t)$ depends on $\Delta_0$ since for larger $\Delta_0$
there is initially a smaller overlap between the solitons and thus they will
approach each other more slowly. By the time the force has changed sign, the
solitons of the molecule with larger $\Delta_0$ have a final separation that is
larger than the one with smaller $\Delta_0$, but both have now to start moving
away from each other, which amounts to different equilibrium points. We have
also checked that the same conclusions are drawn directly from the exact
solution of the GPE.


{\it Other applications. ---} Having an analytic formula for the
interaction potential as a Morse potential, it is appealing to
considering further applications. In particular, we can consider a
classical gas of solitons interacting with the Morse potential in
Eq.~(\ref{eq12}). Here, the thermodynamics is expected to be
enriched by the dependence of the potential on the relative phase
of the solitons.

The equation of state of such an imperfect gas can be derived by a
virial expansion $P/k_B\,T=\rho+B_2\,\rho^2+B_3\,\rho^2+\dots$,
where $\rho$ is the number density and $B_2$ and $B_3$ are the
second and third virial coefficients. The second virial
coefficient is given in terms of the two-body partition function
which reduces to the following integration on the soliton-soliton
potential:
$-(1/2)\int{\left[\exp{(-V_{ss}/k_BT)}-1\right]}\,d\Delta$. Due to
the fact that $V_{ss}$ depends on the initial conditions and in
the partition function all configurations should be taken into
account, an additional integral on all possible initial conditions
should be performed. A simpler but equivalent approach is using
the potential that corresponds to Eq.~(\ref{eq100}), namely
$V_{ss}=-(\alpha^2/2)\,\exp{(-\alpha\,\Delta/4)}\,\cos{\Phi}$,
where $\Phi$ is treated as an independent thermodynamic variable
and the initial conditions are absent from the potential. In the
limit of weak interactions $\alpha\ll k_BT$, a simple expression
can be derived $B_2=-(\pi/8)\alpha^3/(k_BT)^2$, which leads to the
equation of state
\begin{equation}
{P\over k_BT}=\rho-{\pi\over8}\,{\alpha^3\over(k_BT)^2}\,\rho^2.
\end{equation}
Exploring the possibility of a phase transition to a {\it soliton
liquid} requires extending the virial expansion to the third
virial coefficient, which involves three-body processes and hence
requires the knowledge of the three solitons solution, which is
beyond the scope of this Letter.

As another interesting application, we briefly consider a Toda
lattice of solitons, which was successfully used to model the
dynamics of a train of $N$-solitons in external potentials
\cite{new}. The force on the $i$-th soliton at position $z_i$ will
be given by
\begin{equation}
{\ddot z}_i=F_{ss}(\Delta_{i+1})-F_{ss}(\Delta_{i-1}),
\end{equation}
where $\Delta_{i+1}=z_{i+1}-z_i$, $\Delta_{i-1}=z_i-z_{i-1}$, and
$F_{ss}=-dV_{ss}/d\Delta$. The ground state of such a system will
be a series of equidistant solitons with peak intensities
alternating between two slightly different values. The value of
$n_-$ cannot be arbitrarily small since the nearest-neighbor
separation between the solitons in the lattice diverges as
$\log{n_-}$ \cite{usama_pre}. On the other hand, $n_-$ should be
small enough that the solitons are sufficiently separated and that
the coalescence condition $n_+\,g=1$ is avoided \cite{usama_pre}.
In this configuration the net force on any soliton will vanish and
the lattice moves as a rigid object that can be considered as a
``soliton of solitons''. Collective excitations and sound waves
can be considered as the low-energy excitations.

In conclusion, we have shown that a molecule of two bright
solitons can be realized in attractive Bose-Einstein condensates.
This was demonstrated both by numerical simulations and by showing
that the interaction potential between the two solitons is of a
molecular type and takes the form of a Morse potential.
Applications of this potential in the problems of soliton gas and
Toda lattice of solitons were pointed out.

{\it Acknowledgments. ---}
%The first author would like thank the
%Institute for Theoretical Physics at Utrecht University for
%hosting him during a part of this project. 
The authors acknowledge discussions with R.G. Hulet.
This work is supported
by the Stichting voor Fundamenteel Onderzoek der Materie (FOM) and
the Nederlandse Organisatie voor Wetenschaplijk Onderzoek (NWO).


\begin{thebibliography}{99}

\bibitem{randy} D. Dries, S. E. Pollack, J. M. Hitchcock, and R. G. Hulet,
Phys. Rev. A {\bf 82}, 033603 (2010).
%; Pollack, S. E.; Dries, D.; Olson, E.
%J.; Hulet, R. G. American Physical Society, 41st Annual Meeting of the APS
%Division of Atomic, Molecular and Optical Physics {\bf55}, Number 5, abstract
%OPR.40 (2010).

\bibitem{solov} V.I. Karpman and V.V. Solov'ev, Physica D {\bf 3}, 487 (1981).

\bibitem{gordon} J.P. Gordon, Opt. Lett. {\bf 8}, 596 (1983).

\bibitem{anderson} D. Anderson and M. Lisak, Opt. Lett. {\bf 11}, 174 (1986).

%\bibitem{books} A. Hasegawa and Y. Kodama, {\it Solitons in Optical
%Communications} (Oxford University Press, New York, 1995); L.F. Mollenauer and
%J.P. Gordon, {\it Solitons in Optical Fibers} (Acadamic Press, Boston, 2006);
%G.P. Agrawal, {\it Nonlinear Fiber Optics} (Academic Press, San Diego, 2001),
%3rd ed.; N. N. Akhmediev and A. Ankiexicz, {\it Solitons: Nonlinear Pulses and
%Beams} (Chapman and Hall, London, 1997); J.R. Taylor, {\it Optical Solitons -
%Theory and Experiment} (Cambridge University Press, Cambridge, 1992).

\bibitem{books} See e.g., %A. Hasegawa and Y. Kodama, {\it Solitons in Optical
%Communications} (Oxford University Press, New York, 1995);
L.F. Mollenauer and J.P. Gordon, {\it Solitons in Optical Fibers} (Acadamic
Press, Boston, 2006).
%; G.P. Agrawal, {\it Nonlinear Fiber Optics} (Academic
%Press, San Diego, 2001), 3rd ed.; N. N. Akhmediev and A. Ankiexicz, {\it
%Solitons: Nonlinear Pulses and Beams} (Chapman and Hall, London, 1997); J.R.
%Taylor, {\it Optical Solitons - Theory and Experiment} (Cambridge University
%Press, Cambridge, 1992).

\bibitem{mitschke1} M. Stratmann, T. Pagel, and F. Mitschke,
Phys. Rev. Lett. {\bf 95}, 143902 (2005); A. Hause, H. Hartwig, B. Seifert, H.
Stolz, M. B$\rm\ddot o$hm, and F. Mitschke, Phys. Rev. A {\bf 75}, 063836
(2007); A. Hause, H. Hartwig, M. B$\rm\ddot o$hm, and F. Mitschke, Phys. Rev. A
{\bf 78}, 063817 (2008).

\bibitem{perez} L.-C. Crasovan {\it et al.}, Phys. Rev.
E, {\bf67}, 046610 (2003); C. Yin {\it et al.}, arXiv:1003.4617.

\bibitem{santos} R. Nath, P. Pedri, and L. Santos, Phys. Rev. A {\bf76}, 013606
(2007).

\bibitem{expts} K.E. Strecker, {\it et al.}, Nature {\bf417}, 150 (2002); L.
Khaykovich {\it al. et al.}, Science {\bf296}, 1290 (2002); S. L. Cornish, S.
T. Thompson, and C.E. Wieman, Phys. Rev. Lett. {\bf96}, 170401 (2006).

\bibitem{usama_randy}U. Al Khawaja, H.T.C. Stoof, R.G. Hulet,
K.E. Strecker, and G.B. Partridge Phys. Rev. Lett. {\bf89}, 200404 (2002).

\bibitem{toda} M. Toda, {\it Theory of Nonlinear Lattices},
(Springer, Berlin, 1989), 2nd ed.

\bibitem{santosgas} R. Nath, P. Pedri, and L. Santos, Phys. Rev. Lett. {\bf102},
050401 (2009).

\bibitem{usama_pre} U. Al Khawaja, Phys. Rev. E {\bf 81}, 056603 (2010).

\bibitem{new} V.S. Gerdjikov, D.J. Kaup, I.M. Uzunov, and E.G.
Evstatiev, Phys. Rev. Lett. {\bf77}, 3943 (1996); V.S. Gerdjikov,
B.B. Baizakov, M. Salerno, and N.A. Kostov, Phys. Rev. E {\bf73},
046606 (2006).

\end{thebibliography}
\end{document}